\documentclass[aps,prd,showpacs,nofootinbib,notitlepage,12pt]{revtex4-1}
\usepackage{latexsym}
\usepackage{amsbsy,amsmath,amsthm,amsfonts,calc,amssymb,graphicx,accents,mathrsfs,eufrak}
\usepackage{braket,textcomp,upgreek}
\usepackage[percent]{overpic}
\usepackage{color}
\usepackage{psfrag}
\usepackage{enumerate}
\usepackage[colorlinks,citecolor=blue]{hyperref}
\theoremstyle{plain}

\theoremstyle{definition}
\DeclareFontFamily{U}{rsfs}{}         
\DeclareFontShape{U}{rsfs}{m}{n}{<5> rsfs5 <6><7> rsfs7          %
 <8><9><10><10.95><12><14.4><17.28><20.74><24.88> rsfs10}{}     %
\DeclareMathAlphabet{\mathfs}{U}{rsfs}{m}{n}                     %
\definecolor{indiagreen}{rgb}{0.07, 0.53, 0.03}
\def\beq{\begin{eqnarray}}
\def\eeq{\end{eqnarray}}

\def\nn{\nonumber\\}

\def\={\stackrel{\Delta}{=}}

\def\lie{\pounds}

\newcommand*\Laplace{\mathop{}\!\mathbin\bigtriangleup}

\begin{document}
\begin{flushright}
	\footnotesize USTC-ICTS/PCFT-20-09
	\end{flushright}
	\author{Amit Ghosh,}\email{amit.ghosh@saha.ac.in}
	\affiliation{Theory Division, Saha Institute of Nuclear Physics, 1/AF Bidhan 
		Nagar, Kolkata 700064, India}
\title{On Super-Translation transition between Quasi-local Black holes} 
\author{Avirup Ghosh}\email{avirup@ustc.edu.cn}
\affiliation{Interdisciplinary Center for Theoretical Study, University of Science and Technology of China, and 
	Peng Huanwu Center for Fundamental Theory, Hefei, Anhui 230026, China	}
\author{Pritam Nanda,}\email{pritam.nanda@saha.ac.in}
\affiliation{Theory Division, Saha Institute of Nuclear Physics, 1/AF Bidhan 
Nagar, Kolkata 700064, India and HBNI, Mumbai, India}
\begin{abstract}
	We re-explore the symmetries of a weakly isolated horizon (WIH) from the perspective of freedom in the choice of intrinsic data. The supertranslations are realized as additional symmetries. Further, it is shown that all smooth vector fields tangent to the cross-sections are Hamiltonian. We show that joining two WIHs which differ in these Hamiltonians and boundary data, under the action of a supertranslation, necessarily require the inclusion of an intermediate dynamical phase, possibly with the inclusion of a stress energy tensor. This phase of the boundary is non-expanding but not a WIH and invariably leads to a violation of the dominant energy condition. The assumptions made allow us to reconstruct the (classically) pathological stress energy tensor also.
\end{abstract}
\maketitle
\section{Introduction}
Black holes are perhaps the simplest nontrivial solutions of the Einstein equation. The simplicity is precisely encoded in the no-hair conjecture which asserts that black hole solutions of Einstein-Maxwell theory are completely characterised by only three parameters, namely its mass, charge and angular momentum \cite{Israel:1967wq,Carter:1971zc,heusler_1996}, the so called black hole hairs. However, it leads to a number of baffling conclusions about the very nature of black holes. During the formation of a black hole due to the gravitational collapse of matter, whatever the shape of the collapsing object might be, the final state of collapse is only characterised by the object's mass, charge and angular momentum. This is still acceptable owing to the fact that a black hole hides classical information from an asymptotic observer. Its ramifications in the semi-classical context is however of serious concern and gives rise to what is dubbed as the information loss paradox.

The scattering of quantum fields in a classical black hole background was first studied in \cite{Hawking:1974sw}. It was shown that an initial vacuum state prepared at $\mathscr I^-$ would evolve in the black hole geometry to a thermal state at future null infinity $\mathscr I^+.$ Consequently, there is a non-unitary evolution and loss of information. One can imagine this in the context of a collapse process that provides the classical background and a quantum state prepared in the vacuum at $\mathscr I^-$. The out-state at $\mathscr I^+$ being a thermal state, hypothetically implies that the black hole is emitting thermal radiation which causes a decrease in its mass, angular momenta, etc. and may eventually lead to its complete evaporation. Thus as the end state of the collapse and subsequent evaporation one finds the black hole singularity and thermal radiation at $\mathscr I^+$. The information about the collapsing matter is lost. The role of the no- hair conjecture, here, is that the thermal state is characterised only by the nontrivial hairs of the stationary black hole. 

A possible resolution therefore, might be the presence of more hair on the black hole as suggested in \cite{Hawking:2016msc}. It is well-known that mass, angular momentum and electric charge of black holes arise as conserved charges associated with gauge symmetries that become true symmetries in presence of a boundary. Hence one looks for hair by searching for a group of symmetries that are larger than just the group of isometries of a metric. Examples of asymptotically flat space-time at null infinity \cite{Bondi:1962px,Sachs:1962wk, Geroch:1977jn}, asymptotically locally Anti-de-Sitter spacetime \cite{Brown:1986nw}, as well as exploration of near `horizon' symmetries \cite{Carlip:1999cy,Koga:2001vq, Hotta:2000gx} have taught us that this can indeed be the case. The proposal in \cite{Hawking:2016msc}, which is solely motivated from the experiences with asymptotically flat spacetime at null infinity, explores the symmetries on the horizon of a black hole. In the case of $\mathscr I^+$ $(\mathscr I^-)$, the group of symmetries (defined as diffeomorphisms which preserve the fall-off conditions on the metric) becomes infinite dimensional, the so called $\mathcal {BMS^+}$($\mathcal {BMS^-}$), which is a semi-direct product of an infinite dimensional abelian group of supertranslations with the Lorentz group (or its generalisations namely two copies of the Witt algebra \cite{Barnich:2009se} or the algebra of smooth diffeomorphisms on the sphere \cite{Campiglia:2014yka, Campiglia:2015yka} ). Inspite of the similarity of a black hole horizon with $\mathscr I^+ $ or $\mathscr I^-$ this group may not be realised as the symmetries because of the non-affinity of the null generators, especially in the non-extremal case. The Lie-group ideal of supertranslations however turns out to be symmetries that preserve the essential horizon structure. 

 There could be two possible meaning of a supertranslated black hole. It might be the near-horizon supertranslations \cite{Hawking:2016msc} or asymptotic supertranslation at $\mathscr I^+$ and $\mathscr I^-$ acting on global black hole solutions \cite{Hawking:2016sgy,Haco:2018ske}. It is far from whether these two notions are the one and the same, precisely because the extensions of the near-horizon supertranslation generators into the bulk, may not match with the supertranslation generators at $\mathscr I^-$. Here, we will be interested in near horizon supertranslations as opposed to asymptotic supertranslations acting on global black hole solutions.

It has long been hypothesised that asymptotically flat spacetimes related by such supertranslations are inequivalent vacua of the gravitational field \cite{Bondi:1962px}. This gained further ground when it was discovered that the Ward identitites corresponding to supertranslations reproduce the contribution to scattering amplitudes coming from the inclusion of a soft graviton (or a soft photon for asymptotic $U(1)$ gauge transformations) \cite{He:2014laa,Kapec:2015vwa} in Weinberg's soft graviton/photon theorem \cite{PhysRev.140.B516}. The asymptotic states thus differ by the emission of a soft graviton under the action of a supertranslation. This motivates the study of supertranslations near the black hole horizon as well. The proposal in \cite{Hawking:2016msc} has attracted a plethora of work that explores the symmetries near the black hole horizon \cite{Donnay:2016ejv,Donnay:2015abr,Averin:2016ybl,Afshar:2016wfy,Setare:2016jba,Mao:2016pwq}. These works mostly aimed at finding extended group of symmetries, larger than the just the supertranslations \cite{Donnay:2016ejv,Donnay:2015abr,Afshar:2016wfy,Setare:2016jba}, and also checked whether the entropy of a black hole can be accounted for by counting the dimension of a representation of such an extended symmetry group \cite{Carlip:2017xne}.

Our aim in this paper is not to look for extended symmetries but to study transitions between supertranslated quasi-local black holes akin to the vacuum to vacuum transitions studied for the case of asymptotic symmetries at null infinity \cite{Compere:2018ylh}. It is well-known that if there is a non-stationary epoch between two stationary epochs then the natural frames of the two stationary epochs are related by a supertranslation and a boost \cite{Strominger:2014pwa, Hollands:2016oma}. In the case of a black hole horizon or more specifically a quasi-local black hole, described by the existence of an apparent horizon, the stationary regions are described by boundaries called (weakly) Isolated horizons \cite{Ashtekar:1998sp}, while the transition between two Isolated horizons is effectively captured by a Dynamical horizon \cite{Ashtekar:2003hk}. In a general transition the metric on the cross-section may evolve arbitrarly (depending on the flux) in contrast to null infinity where they are just the round metric on the sphere \footnote{The metric can be conformal to the round sphere or any metric on $S^2$, depending on the kind of boundary condition one imposes. This leads to the other two asymptotic symmetry groups where it is a semidirect product of the Witt algebra or Diff($S^2$) respectively, with that of the Lie idela of supertranslations. But dynamical transitions are studied in the so-called Bondi frame where the metric is indeed just the round metric on the sphere.}. Moreover, there is no accepted definition of surface gravity in the dynamical phase, as compared to the Bondi mass at null infinity. These directly appear in the expressions for supertranslation charges. Hence there is no direct generalisation of the evolution of supertranslation charges in the dynamical regime, making it diffcult to compare them across the dynamical epoch. 

The relation between preferred foliations in late and early stationary regions of an evolving black has however been argued to be related by supertranslations in a framework where dynamical evolution of black holes is described by an expanding and shearing null congruence namely the event horizon \cite{Rahman:2019bmk}. In the foregoing discussion we will not deal with such  general evolutions but try to preserve all boundary data that are essentially preserved by the action of supertranslations in the stationary era. Hence in our case the dynamical phase would not truly be a Dynamical horizon but will continue to be a non-expanding horizon. But, as we will see, it will not obey the weakly condition which necessarily has to be imposed so that the metric in the neighborhood is stationary to linear order (in distance from the horizon).

The paper is organised as follows. In the first few sections we recall the definition of an isolated horizon and try to discuss the intrinsic freedom in the choice of the boundary data. The discussion closely follows \cite{Ashtekar:2004gp} but our argument is essentially in the reversed order. In recent times there has been discussions on symmetries of intrinsic geometry of non-expanding horizons \cite{Sousa:2017auc}, null shells \cite{Blau:2015nee, Bhattacharjee:2017gkh} and general null surfaces \cite{Chandrasekaran:2018aop}. While, we recover some (appropriate to our choice of boundary conditions) of these results, our argument differs from these. We then calculate conserved charges for the horizon supertranslations and show that arbitrary smooth diffeomorphisms on the cross-sections are also Hamiltonian (they are not however symmetries with the kind of bounadry conditions we choose). We then go on to find the action of the supertranslations on various geometric quantities on the horizon which are required to reconstruct the metric in the neighborhood. Guided by these changes we construct a region where such changes are induced by the inclusion of some matter flux. We partially reconstruct the non-zero stress energy tensor components required to sustain such boundary conditions.   
\section{Weakly Isolated horizon}
In this section, we revisit the definition of an Isolated Horizon (IH) \cite{Ashtekar:1998sp, Ashtekar:2000sz, Ashtekar:2000hw, Ashtekar:2001is,Ashtekar:2001jb}. Let $\mathcal{M}$ be 
a smooth $4$ dimensional manifold equipped
with a metric $g$ of signature $(-,+,+,+)$. Let $\Delta$  be a null hypersurface 
 in $\mathcal M$, `$l$'
being it's future directed null normal. Together with $l$ one can define a set of null bases, $(l\,\, n\,\, m\, \bar{m})$, adapted to $\Delta$. They satisfy the usual cross normalisation relations. Let `$h$' be the degenerate metric on the 
hypersurface, and `$q$' the metric on the cross-sections $S_{\Delta}$ of the horizon.  Let us denote the coordinates on $\Delta$ as $\chi^a$ and those on the cross-sections $S_{\Delta}$ as $\tau^A$. We will use the notations $h_{ab}:=h(\partial_a,\partial_b):=g(\partial_a,\partial_b)$ and $q_{AB}:=q(\partial_A,\partial_B):=g(\partial_A,\partial_B)=q(\partial_A,\partial_B)$. The expansion $\theta_{l}$ of the null normal is given in terms of the extrinsic curvature of $S_{\Delta}$ as an embedding in $\mathcal M$ viz.
$\tilde q^{AB}g(\partial_A,\nabla_{\partial_B}l)$. In terms of the Newman Penrose co-effecients,
$\theta_l:=-2\rho$ (see appendix \ref{NP} for details). One defines an equivalance class of 
null normals $[l]$ such that 
two null normals $l$ and $l'$ will be said to belong to the same equivalance 
class if $l'=cl$ where $c$ is a constant on $\Delta$.

{\it{Definition}}: A null hypersurface $\Delta$ of $\mathcal{M}$ is said to be an Isolated horizon if the following conditions hold.:
\begin{enumerate}
\item $\Delta$ is topologically  $S^2\times R$ and null.
\item The expansion $\rho$ of $l$ vanishes on $\Delta$ for any null normal $l$ in $[l]$.
\item All equations of motion hold on $\Delta$ and the stress-
energy tensor $T$ on $\Delta$ is such that $-T^a_{~b}~l^b$
is future directed and causal.
\end{enumerate}
The null hypersurface $\Delta$ is equipped with a one form $\omega_{a}:=-g(n,\nabla_{\partial_a}l)$ on $\Delta$. When $l,~n$ are viewed as vectors normal to the cross-sections $S_{\Delta}$ of $\Delta$, then the pull back of $\omega$ onto $S_{\Delta}$ is nothing but the connection in the normal bundle of $S_{\Delta}$ through the Weingarten map. The IH is called {\bf weakly} if the condition $[\lie_l,\mathbb D] W=0$ holds, where $\mathbb D$ is the intrinsic connection on $\Delta$ and $W\in T\Delta$. 
\section{Conformal transformation and action on the boundary data}
 The first order structure on $\Delta$ is the degenerate metric and the equivalence class of null normals $[l]$. Let us first choose a representative element of the equivalence class and try to define the second order structures with it. One can then find out how the second order structure varies over the equivalence class. The second order structure consists of the connection on $\Delta$. The following conditions will be assumed to be satisfied by $\mathbb D$ on $\Delta$. 1) ~The connection annihilates the degenerate metric i.e $\mathbb D_ch_{ab}=0$ and 2)~The connection is torsion-free.
Let us now denote the connection in terms of its components $\it\Upgamma^c_{ab}\partial_c:=\mathbb D_{\partial_a}\partial_b$. The second condition ensures that the coeffecients $\Upgamma^c_{ab}$ are symmetric in the lower two indices. 
Since $q_{ab}$ is non-invertible, the two conditions do not uniquely fix $\Upgamma$. Let us denote by $h^{ab}$, the inverse of $h_{ab}$ in a non degerate subspace of $h$. In particular this can be just the inverse of the metric $q$ on $S_{\Delta}$. It is one of the many inverses of $h_{ab}$. One can then write a representative element $\Upgamma$ as,
\begin{gather}
\Upgamma^c_{ab}=\frac{h^{cd}}{2}\bigg(\partial_ah_{db}+\partial_{b}h_{da}-\partial_dh_{ab}\bigg)+l^{(c}V^{d)}\bigg(\partial_ah_{db}+\partial_{b}h_{da}-\partial_dh_{ab}\bigg),
\end{gather}
where $V:=V^a\partial_a$ is any vector tangent to $\Delta$. Due to the fact that $l$ is a Killing vector of $h_{ab}$ (which follows from the expansion free and shear free nature of $\Delta$ ), the expression reduces to,
\begin{gather}
\Upgamma^{c}_{ab}=\accentset{\circ}{\Upgamma}^{c}_{ab}+\frac{1}{2}l^c\lie_Vh_{ab}+\frac{1}{2}l^c\partial_{(a}V_{b)}
\end{gather}
It follows that for any covariant vector $\underline\alpha$ on $\Delta$ such that $l.\underline\alpha=0$, the ambiguous part vanishes. The covariant derivative thus has a unique action on such vectors. We will consider the action of $\mathbb D$ on $\underline n$ later. For now let us consider its action on a contravariant vector $W$ tangential to $\Delta$. Note that the contraction of the amibiguous piece with $\underline{m}$ and $\underline{\bar m}$ is zero, thus determining the action of $\mathbb D$ on $W$ uniquely in these directions. This also follows from the fact that the action of $\mathbb D$ on $\underline{m}$ and $\underline{\bar m}$ are known uniquely. The action of $\mathbb D$ on $W$, in the directions $\underline n$ is however not known uniquely. In fact the action on $l$ is completely arbitrary as the unambiguous piece does not contribute in this case. It is therefore of the form,
\begin{gather}
\mathbb D_{\partial_a}l=\upomega_a~l,
\end{gather}
 where $\upomega$ is completely arbitrary. Given a fixed bulk spacetime, the pull back of the bulk connection of course defines `a' $\omega$, introduced in the previous section, by $\nabla_{\partial_a}l=\omega_a~l$. Therefore the intrinsic connection defines an equivalence class $[\upomega]$ such that $\omega\in[\upomega]$. This equivalence class can be further restricted so that physical quantities defined from every element are same. The weakly condition is essentially such a restriction, which along with the Dominant energy condition implies that, 
\begin{gather}
d(l.\omega)=0
\end{gather}
thus restricting the equivalence class to $\upomega_1\sim\upomega_2$ if $\upomega_1-\upomega_2=\underline\beta$, such that, $l.\underline{\beta}=constant$. This constant can be set to zero by demanding that the surface gravity $\kappa$ does not vary over the equivalence class. Further, recall that the exterior derivative of $\omega$ is used to define quasi-local multipole moments of ${\Delta}$ \cite{Ashtekar:2004gp}. In order that there is no ambiguity in this definition, we must also ensure that $\underline \beta$ be at most of the form, $d\beta$ (say), for some function $\beta$ such that $\lie_l\beta=0$. We would like to include this restricted ambiguity in the definition of symmetries of an IH. We will see that it is precisely this freedom that the supertranslations on $\Delta$ generate. 

It turns out that the symmetries can be effectively studied using conformal transformation, in analogy with $\mathscr I^+$. The initial choice of conformal frame is already fixed on an IH. The conformal transformation that we are going to study are analogous to the residual conformal freedom at $\mathscr I^+$. A conformal transformation of the metric amounts to a conformal transformation of the 
degerate metric on $\Delta$.
Under such a transformation $g\rightarrow \Omega^2\,g$ and $l\rightarrow \Omega^{-1}l,~\underline{l}\rightarrow \Omega~ \underline{l},~n\rightarrow \Omega^{-1} n,~\underline{n}\rightarrow \Omega \underline{n},~m\rightarrow \Omega^{-1}m,~\underline{m}\rightarrow \Omega \underline{m}$. The new derivative 
operator, compatible with the transformed metric is such
%
%
%
%
that the Newman Penrose scalars transform in the following way,
\begin{eqnarray}
\tilde{\rho}\=\Omega^{-1}~(\rho-\lie_l{\log\Omega}),~~\tilde{\sigma}\=\Omega^{-1}~\sigma,~~\tilde\kappa_{NP}\=\Omega^{-1}~\kappa_{NP}
\end{eqnarray}
where, $\rho=-g(\bar m,\nabla_m~l)$ and 
$\sigma=-g(\bar m,\nabla_{\bar m}~l)$. Therefore, all conformal transformations obeying $\lie_l\Omega\stackrel{\Delta}{=}0$ keep the isolated horizon boundary conditions invariant. One must also ensure that the zeroth law for black holes, that is the extra conditions imposed to make it a WIH, remains intact. To do this one must check how $[\upomega]$ transforms. A representative element transforms as
\begin{gather}
\omega\rightarrow \Omega^{-1} ~ \omega +d\log\Omega,\label{comega}
\end{gather}
In order that the zeroth law remains valid one therefore must choose $\Omega$  to be a constant. Another quantity of interest is the action of the covariant derivative on $\underline{n}$. This quantity is also not unambiguously defined as discussed previously. It turns out that the pullback of covariant derivative of $\underline n$ onto $S_{\Delta}$ is nothing but the extrinsic curvature of $S_{\Delta}$ along the $n$ direction i.e $g(\partial_A,\mathbb D_{\partial_B}n)$. Therefore one can write its evolution equation as (eq. (\ref{ETEC})),
\begin{gather}\label{WTEC}
\lie_{l}g\big(n,K(\partial_A,\partial_B)\big)=-\kappa~g\big(n,K(\partial_A,\partial_B)\big)+\frac{1}{2}~^2\mathcal R(\partial_A,\partial_B)-\frac{1}{2}R(\partial_A,\partial_B)-\mathcal D_{(A}\omega_{B)}-\omega_{(A}\omega_{B)},
\end{gather}
where $\omega_A=-g(n,\nabla_{\partial_A}l)$, which we call the rotation one-form, is the pull-back of $\omega$ onto the cross sections. The weakly condition now implies that the left hand side is zero \cite{Ashtekar:2004gp}, thus determining $K^{(n)}(\partial_A,\partial_B)$ locally (in time) and uniquely in terms of the intrinsic data. Note that this condition also implies that the next to leading order metric $\accentset{(1)}{q}_{AB}$ (say) (in the neighbourhood of $\Delta$) is stationary. A further derivative on the expression for $K^{(n)}(\partial_A,\partial_B)$ would then imply that $R(\partial_A,\partial_B)$ is independent of time. 
\section{Symmetries}
We will now try to find the vector fields that generate the conformal diffeomorphisms that preserve the conditions upto the equivalence class. As mentioned before, for the case of a nonextremal horizon, restricting the conformal transformations to those for which $d\Omega\stackrel{\Delta}{=}0$, preserves all the boundary conditions as well as the zeroth law. One can therefore consider this to be a symmetry of the  nonextremal isolated horizon boundary conditions and look for infinitesimal versions of it viz.,
\begin{equation}
\lie_\xi l=c~l,~~\lie_\xi ~q=-2c~q,
\end{equation}
where $c$ is a constant on $\Delta$. It is easy to see that the generators form an algebra isomorphic to the semi-direct product of supertranslation with the homothetic Killing vectors of $q$. The generators are given by,
\begin{gather}
\xi=f~l~~~{\text {such that}}~~~~~~~~~\lie_lf=0\nn
\xi=a~l+\eta~~~~{\text {such that}}~~~~~~~g=b~v,~~\lie_\eta~ q=-2b~q,\label{sym}
\end{gather}
where $f$ is a function of the coordinates on the cross-sections, $b$ is a constant $v$ is the parameter along $l$. Let us now evaluate the action of these vector fields on $\kappa$ and $\omega$ so as to check that they indeed generate the correct transformations. Let us start with the action of the supertranslations. The supertranslations are given by $c=0$ which corresponds to the $\Omega=1$ case. The Lie-derivation acting on $\omega$ is given as,
\begin{gather}
\lie_{fl}\omega\stackrel{\Delta}{=}df~(l.\omega),
\end{gather}
which is consistent with the ambiguous piece in eq.(\ref{comega}). Hence it is clear that the supertranslation essentially generate these ambiguities. Let us now go over to the Homotheties. The transformation would be of the form,
\begin{gather}
\lie_{\xi}\omega=h~dv~(l.\omega)+\lie_{\eta}\omega.
\end{gather}
To calculate the term $\lie_\eta\omega$ recall the expression for $\Gamma$. Under a homothetic transformation the unambiguous part does not transform, only the ambiguous part transforms by a multiplicative constant. Hence the condition is satisfied. Here, we will however not be concerned with homotheties as on a complete and connected Riemannian manifold all homotheties reduce to isometries \cite{kobayashi1996foundations}. The transformation within the equivalence class of null normals $[cl]$ is generated by local boosts. These can be studied independently, but would be irrelevant for our discussion.
\section{Symplectic structure and charges}
The variation of a Lagrangian is on-shell of the form $\delta 
L=d\Theta(\delta)$. $\Theta$ 
is called the symplectic potential. The symplectic structure,$\Omega (\delta_{1}, \,\delta_{2})$, on the space of solutions is then defined as,
\begin{equation}
\Omega(\delta_1,\delta_2)=\int_{M}J(\delta_1,\delta_2),
\end{equation}
where $J(\delta_1,\delta_2)= 
\delta_1\Theta(\delta_2)-\delta_2\Theta(\delta_1)-\Theta(\delta_1\delta_2-\delta_2\delta_1)$ and $M$ is a space-like hypersurface. The symplectic structure is closed on shell 
.ie, $dJ=0$. On integration over a region of spacetime, bounded by spacelike surfaces $M_1$, $M_2$ and boundary $B$ ($M_1\cup M_2\cup 
B$) this gives,
\begin{equation}
\int_{M_1}J-\int_{M_2}J~+~\int_{B}J=0,
\end{equation}
Depending on the boundary conditions, the third term may vanish, in which case the bulk symplectic structure is independent of 
the choice of hypersurface or it may be  exact, i.e $\int_{B}J=\int_{B}dj $
implying that,
\begin{equation}
\Omega(\delta_{1}, \,\delta_{2})= \int_MJ-\int_{S_B}j
\end{equation}
is the hypersurface independent symplectic structure. Here, $S_B$ is the 2-surface $M\cap B$ and $j(\delta_1,\delta_2)$ is called the boundary symplectic current. For the case of first order gravity in the Palatini formulation,
\begin{gather}
L=\frac{1}{16\pi G} \Sigma_{IJ}\wedge F^{IJ}\label{lagrangian}
\end{gather}
where $\Sigma_{IJ}=\epsilon_{IJKL}e^k\wedge e^L$, $F^{IJ}=d\omega^{IJ}+\omega^{IK}\wedge\omega_{K}~^J$ and $\omega ^{IJ}$ is the $SO(3,1)$ connection defined as, $de^I=de^I+\omega^I~_J{\wedge e^J}=0$. The symplectic potential in this case is 
given by, $16\pi G\Theta(\delta)=-\Sigma^{I\!J}\wedge \delta A_{I\!J}$ while the 
symplectic current as,
\begin{equation}\label{symplectic_current1}
J_G(\delta_1,\delta_2)=-\frac{1}{8\pi 
	G}\,\delta_{[1}\Sigma^{IJ}\wedge~\delta_{2]}A_{IJ}
\end{equation}
The pull back of the above expression, eqn. \eqref{symplectic_current1} to $\Delta$ can be shown to be exact \cite{Ashtekar:1999yj}, thus defining the boundary symplectic current as,
%
%
%
\beq
\underleftarrow{J_G}(\delta_1,\delta_2)&\stackrel{\Delta}{=}&-\frac{1}{4\pi G}
d\left[\left(\delta_{[1}~^2{\bf\epsilon}~\delta_{2]}\psi\right)\right].
\eeq
where $\psi$ is a potential for the surface gravity, $\kappa$, defined as $\lie_l\psi=\kappa$. We shall use this to calculate the charges for the supertranslation generators and further show that all smooth vector fields tangent to the cross-sections are Hamiltonian.
\subsection*{Hamiltonian Charges and their Algebra}
We can now proceed to study the generators of the symmetries ( eq. (\ref{sym})) on phase space. Given a vector $X$ on spacetime, its action on the dynamical variables is naturally given by the lie derivative $\lie_X$. For any such vector field one can show, using Einstein's equation, that,  
\begin{eqnarray}
\Omega(\delta,\delta_X)&=&-\frac{1}{16\pi 
	G}\int_{S_{\Delta}}\left[(X.A_{IJ})\delta\Sigma^{IJ}-(X.\Sigma^{IJ})\wedge 
\delta A_{IJ}\right]+\frac{1}{8\pi 
	G}\int_{S_{\Delta}}\left(\delta~^2{\bf\epsilon}~\delta_{X}\psi-\delta_X~^2
{\bf\epsilon}~\delta\psi\right)\nn
\end{eqnarray}
%
Let us now consider the supertranslations i.e the case for which $X=fl$, where $\lie_lf=0$.
\begin{gather}
\Omega(\delta,\delta_X)=-\frac{1}{16\pi G}\int_{S_{\Delta}}f\kappa~\delta~^2\varepsilon.
\end{gather}
The contribution from the boundary symplectic structure is decided by the action of $X$ on $\psi$. To find this out, consider the defining relation for $\psi$ i.e  $\lie_l\psi=\kappa$. Now the action of $\delta_X$ on both sides of the above relation gives $\delta_X\lie_l\psi=\lie_l\delta_X\psi=\delta_X\kappa=0$. It follows that if $\delta_X\psi$ is set to zero at the initial slice, it remains so on every slice. The action of $X$ on $~^2\varepsilon$ can be obtained by the action of $\lie_X$ on it. Consequently, the contribution from the bounday symplectic structure is zero. It is hamiltonian only if $\kappa$ does not vary over phase space. 
%
%
%

We will next consider all smooth vector fields on $S_{\Delta}$. Though they are not symmetries, they will be shown to be Hamiltonian. Consider an aribitrary vector field tangent to the cross sections $S_{\Delta}$ and commuting with $l$. The expression for the symplectic structure then gives,
\begin{gather}
\Omega(\delta,\delta_\eta)=-\frac{1}{8\pi G}\int_{S_{\Delta}}\eta.\omega~\delta~^2\epsilon-\eta.~^2\epsilon ~\delta\omega=-\frac{1}{8\pi G}\int_{S_{\Delta}}\delta(\eta.\omega~^2\epsilon)
\end{gather}
The action of $\eta$ on $\psi_l$ can be found by a procedure similar to the one employed for supertranslations. Since $l$ commutes with these vector fields it follows that one can set $\delta_\eta\psi_l=0$ as in the case of supertranslations. The action of $\eta$ on $~^2\epsilon$ gives $(div~\eta)~^2\epsilon$. Hence if the vector fields are divergence free then the boundary symplectic structure completely vanishes. However note that for the supertranslations to be Hamiltonian we required $\kappa$ to be a constant on phase space i.e $\delta \kappa=0$. We can as well set $\delta\psi=0$ on $\Delta$ thus making all such $\eta$ Hamitonian.

The Poisson bracket of the supertranslation generators with $\mathcal H_\eta$ is then given by,
\begin{gather}\label{superchange}
\Omega(\delta_{fl},\delta_\eta)=\frac{1}{8\pi G}\int_{S_{\Delta}} \kappa~ df\wedge\eta.~^2\epsilon=\frac{1}{8\pi G}\int_{S_{\Delta}} \kappa~ \eta.df~^2\epsilon
\end{gather}
 These represent the change of the charge $H_{\eta}$ under a supertranslation. The interesting point to note here is that the change is zero if $\eta$ is divergence free. Hence the supertranslations do not change the multipolar structure of the IH as was anticipated before. We will be equating these changes across a phase, not satisfying the weakly condition, and try to find out the content of the Stress energy tensor during the transition.
\section{Action of super-translations on the horizon data}
Here we will look at the action of super-translations on the horizon data. Let us assume that there is a parameter $s$ on `phase space' which acts as an affine parameter for the generators of super-translations. This is necessarily not the affine parameter of $fl$ on space-time and  in general is a functional of $\omega,~f, ~^2\epsilon$. Let us now consider the change of $\mathcal H_\eta$ along this parameter.
\begin{gather}
\frac{\delta\mathcal H_\eta}{\delta s}=\frac{1}{8\pi G}\int_{S_{\Delta}} \kappa~ \eta.df~^2\epsilon
\end{gather}
It is clear that the right hand side of the above equation is independent of $s$ as $\{\{\mathcal H_\eta,\mathcal H_{fl}\},\mathcal H_{fl}\}=0$. Therefore the total change is 
\begin{gather}
\Delta\mathcal H_{\eta}=\frac{\Delta s}{8\pi G}\int_{S_{\Delta}} \kappa~ \eta.df~^2\epsilon
\end{gather}
Similary for $\omega$ we have, 
\begin{gather}
\frac{\delta}{\delta s}\omega_A=\kappa~ \partial_Af,~~~~~~~~~~\omega_A=\omega_A^0+\kappa ~\partial_A(fs)\label{superchange2}
\end{gather}
Note that these relations also represent the changes under finite supertranslations. Let us therefore denote $fs$ as a general function $\mathcal F$. To find the transfomation of $K^{(n)}_{AB}$ the expression for $\omega_A$, eq. (\ref{superchange2}), is substituted in the expression for $K^{(n)}_{AB}$, obtained after imposing the weakly condition eq. (\ref{WTEC}), resulting in,
\begin{gather}
K^{(n)}_{AB}=-\frac{1}{\kappa}\bigg[-\frac{1}{2}~^2\mathcal R(\partial_A,\partial_B)+\frac{1}{2}R(\partial_A,\partial_B)+\mathcal D_{(A}\omega^0_{B)}+\omega^0_{(A}\omega^0_{B)}\nn
+\kappa~\mathcal D_{(A}\mathcal D_{B)}\mathcal F+\kappa~\mathcal D_{(A}\mathcal F~\omega^0_{B)}+\kappa~\omega^0_{(A}\mathcal D_{B)}\mathcal F+\mathcal \kappa^2~\mathcal D_{(A}\mathcal F~\mathcal D_{B)}\mathcal F\bigg]\label{ECT}
\end{gather}
The last term in the above expression is of quadratic order and can be ignored when one considers linearised supertranslations. As we will see later, the next to leading order metric on the cross sections in the neighborhood (~$\accentset{(1)}{q}_{Ab}$~) can be obtained just by multiplying the radial coordinate to the above expression. Hence the transformation law should be such that the transformation of $\accentset{(1)}{q}_{AB}$ is consistent with those obtained previously in the literature. It is reassuring that this expression matches with the supertranslated metric in \cite{Donnay:2016ejv}, with the time derivative term removed. This is precisely because of the weakly condition on the un-supertranslated metric, as discussed before.

Let us now try to recover these from the perspective of diffeomorphisms on $\Delta$. Let us denote the supertranslations as a map $\psi:\tilde\Delta\rightarrow\Delta$. Calculating the transformed $\omega$ then amounts to obtaining the pull-back connection under the map $\psi$. First note that $\psi^*n=n-\kappa ~d\mathcal F$. Recall that the definition of the Pull-back connection is,
\begin{gather}
\psi_*\bigg((\psi^*\mathbb D)_XY\bigg)=\mathbb D_{\psi_*X}\psi_*Y
\end{gather}
and note that $\psi_*l=l$ and $\psi_*\partial_A=\partial_A\mathcal F~l+\partial_A$. Thus we have,
\begin{gather}
\psi_*(\tilde\kappa~l)=\kappa~ l\\
\psi_*(\tilde\omega_A ~l)=(\omega_A+\kappa~\partial_A\mathcal F)l.
\end{gather}
Let us now consider the extrinsic curvature $K^{(n)}(\partial_A,\partial_B)$.
\begin{gather}
\psi_*\bigg((\psi^*\mathbb D)_{\partial_A}\partial_B\bigg)=\mathbb D_{\psi_*\partial_A}\psi_*\partial_B=\mathbb D_{\partial_A}\partial_B+\mathcal D_A\mathcal F~\mathbb D_{\partial_B}l+\mathcal D_B\mathcal F~\mathbb D_{\partial_A}l+\mathcal D_A\mathcal F~\mathcal D_B\mathcal F~\mathbb D_l~l\nn
+\mathcal D_A\mathcal D_B\mathcal F~l,
\end{gather}
where $\mathcal D$ is the covariant derivative compatible with $q$, the metric on the cross-sections. This correctly reproduces the transformation eq. (\ref{ECT}) on contracting both sides by $n-\kappa~d\mathcal F$.
\section{Inducing a Super-Translation}
\begin{figure}[h]
	\begin{minipage}[h]{0.4\paperwidth}
		\begin{overpic}[width=\linewidth]{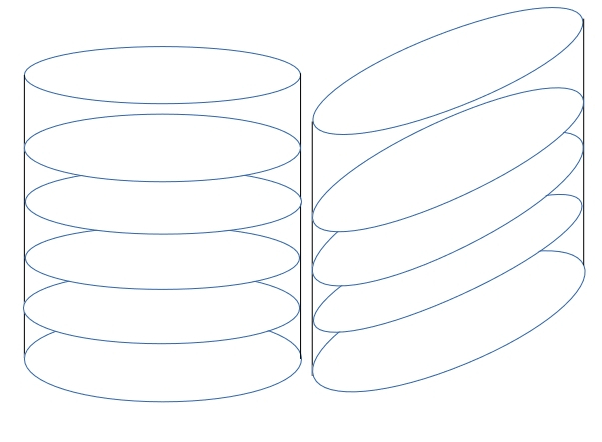}
			\put(70,70){\small$\Delta_2$}
			\put(25,70){\small$\Delta_1$}
		\end{overpic}
		\caption{Representative of two IH's $\Delta_1$ and $\Delta_2$ that differ by a supertranslation hair.}
	\end{minipage}
	\begin{minipage}[h]{0.3\paperwidth}
		\begin{overpic}[width=\linewidth]{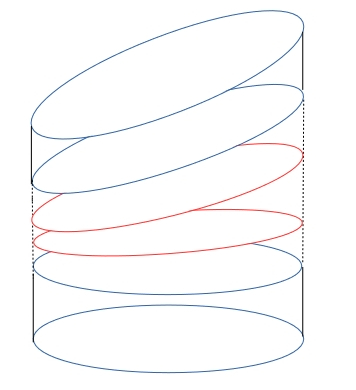}
			\put(43,80){\small$\Delta_2$}
			\put(43,15){\small$\Delta_1$}
			\put(43,39){\small$\mathcal H$}
			\put(80,30){\small$\Sigma_2$}
			\put(80,75){\small$\Sigma_2'$}
			\put(80,13){\small$\Sigma_1$}
			\put(80,94){\small$\Sigma_3$}
		\end{overpic}
		\caption{Representative of the gradual change of foliation in the dynamical phase.}
	\end{minipage}
\end{figure}
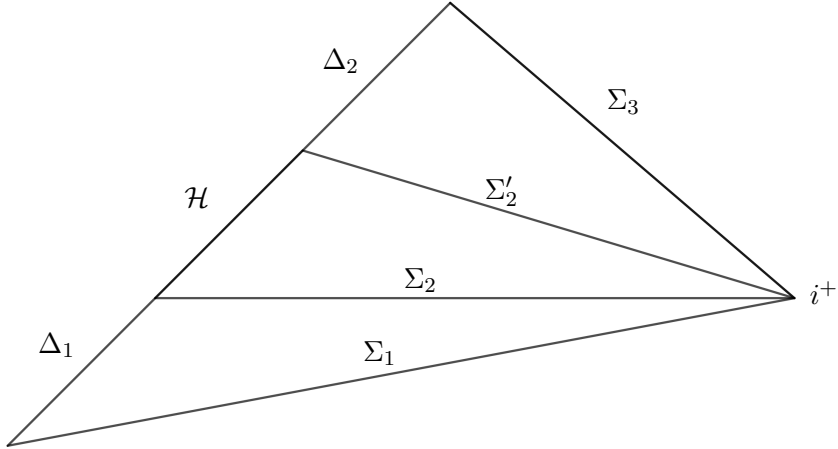
\begin{figure}[h]
	\begin{minipage}[h]{0.5\paperwidth}
		\begin{overpic}[width=\linewidth]{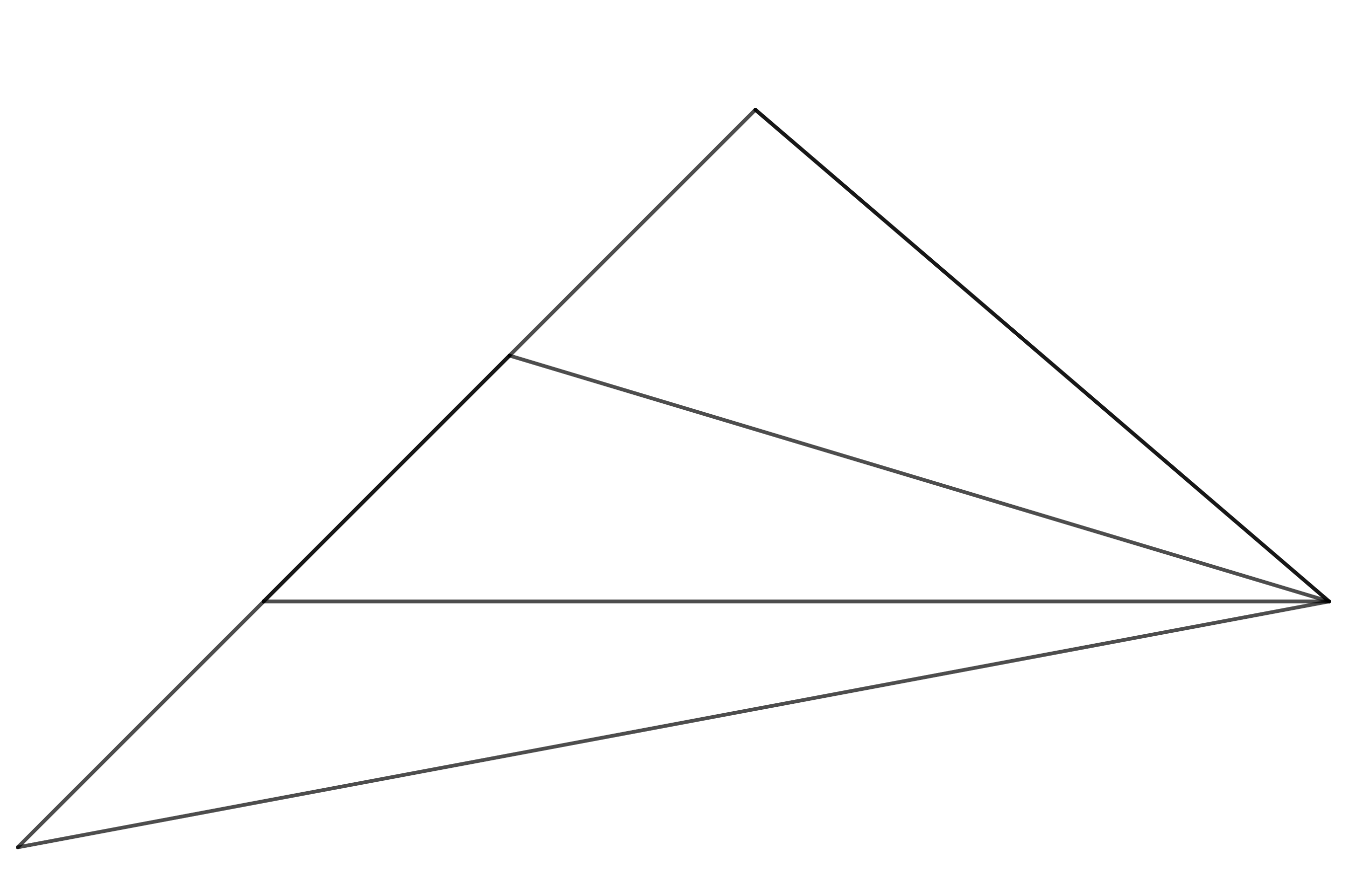}
			\put(40,50){\small$\Delta_2$}
			\put(5,15){\small$\Delta_1$}
			\put(23,33){\small$\mathcal H$}
			\put(50,23){\small$\Sigma_2$}
			\put(60,34){\small$\Sigma_2'$}
			\put(45,14){\small$\Sigma_1$}
			\put(75,45){\small$\Sigma_3$}
			\put(100,21){\small$i^+$}
		\end{overpic}
		\caption{The contruction to implant or excite such a hair in a dynamical transition between the two stationary states.}
	\end{minipage}
\end{figure}
Having obtained an expression for the change in the boundary data under the action of a supertranslation, we would like to construct a dynamical situation where such a change is induced. In general one would think that such a change can be induced when an isolated horizon is transitioning from one equillibrium state to another through a Dynamical horizon. The construction however should be such that 1) the change in $\mathcal H_\eta$ or $\omega_A$ is correctly reproduced, 2) there is no change in the area two-form on the cross-sections (as the supertranslations Lie-drag the area two-form), 3) the metric on the cross-sections is Lie-dragged. Let us now recall the definition of a dynamical horizon. A dynamical horizon $\mathcal H$ is defined by a spacelike surface foliated by marginally trapped surfaces (MTS's), given by the conditions $\theta_l=0$ ($l$ is outgoing and future directed). Thus each cross-section of $\mathcal H$ is an MTS and there is an evolution vector $X$ that maps one such cross section $S_{\mathcal H}$ of $\mathcal H$ to another during the evolution. Let us suppose that we have smoothly extended the null frame in the equilibrium regions to construct a null frame on each of the $S_{\mathcal H}$'s. The evolution vector can then be taken to be $X=\alpha~l-\beta~n$ \cite{Booth:2006bn}. We can also construct a timelike vector $\tau=\alpha~l+\beta~n$ which is orthogonal to $X$.  $\alpha$ and $\beta$ are required to satisfy the following equation on $S_{\mathcal H}$ such that it is mapped into another MTS,
\begin{gather}
\Laplace^{S_{\mathcal H}}\beta-2\omega^A\partial_A\beta-\beta \mathcal D_A\omega^A+\beta\omega^A\omega_A-\beta\bigg(\frac{^2\mathcal R}{2}-T(l,n)\bigg)-\alpha\bigg(T(l,l)+K^{(l)}_{AB}K^{(l) AB}\bigg)=0,\label{deviation}
\end{gather}
where $\Laplace^{S_{\mathcal H}}$ and $\mathcal D$ are the Laplace operator and the covariant derivative on $S_{\mathcal H}$. $K^{(l)}_{AB}$ is the extrinsic curvature of $l$, $^2\mathcal R$ is the Ricci scalar of $S_{\mathcal H}$ and $\tau^A$ are coordinates on $S_{\mathcal H}$. This expression has been obtained several times in the literature \cite{Booth:2006bn} e.g. The change in the area form along the dynamical horizon is given by,
\begin{equation}
\lie_X~^2\epsilon=\bigg(\alpha\theta^{(l)}-\beta\theta^{(n)}\bigg)~^2\epsilon
\end{equation}
The total change of area must be zero because of our requirement. One can think of various situations by which this can happen. For example, the area might first increase and then decrease or vice-versa, such that the total area change is zero. Both these processes however violate the area increase law. Consequently, one must think of a situation where area remains constant. Clearly this implies that $\beta$ must be zero, so $\mathcal H$ must be non-expanding. Note that we still haven't imposed the weakly condition. Hence $\mathcal H$ may not be an IH. Also note that eq.(\ref{deviation}) does not impose any condition on $\alpha$.

Before discussing the evolution of the rotation one form let us first discuss the ambiguities in choosing a foliation of $\mathcal H$. One could have added an arbitrary vector on $\mathcal S_{\mathcal H}$ to $X$. But the condition that the metric on $\mathcal S_{\mathcal H}$ be Lie-dragged then necessitates that this be a Killing vector. In the absence of which the evolution vector can be taken to be just $l$. Either of the IH's when considered globally by itself has a preferred foliation defined by $\mathcal D^A\beta_A=0$ \cite{Ashtekar:2000sz}. But in the current situation, where they are being joined, preferred foliations are necessarily not mapped into preferred foliation as that would mean that the supertranslation is trivial. The choice of $l$ as the evolution vector here means that neither of the two IH's are preferrably foliated. One just chooses an arbitrary foliation by a null vector, throughout $\Delta_1\cup\mathcal H\cup\Delta_2$. The changes are imposed through the evolution of $\omega_A$. We will further assume that $l$ has constant accelaration, as this choice can always be made on a non-expanding horizon \cite{Lewandowski:2006mx}. We will discuss its consequences shortly. 

 Let us now find the condition imposed by enforcing the change in $\mathcal H_\eta$ in the dynamical process. In the subsequent calculation we will only assume the non-expanding condition on $\mathcal H$. We will not impose the `weakly' condition. We will also not be assuming any gauge conditions right now. We conclude that the total change in $H_\eta$ can be written as (appendix \ref{ERO}),
\begin{gather}
\Delta H_\eta=\frac{1}{8\pi G}\int_{v_1}^{v_2} dv\int_{S_{\mathcal H_v}}[\partial_A\kappa(v,\tau^A)+R(\eta,l)]~^2\epsilon\nn
=\int_{v_1}^{v_2} dv\int_{S_{\mathcal H_v}}[\partial_A\kappa(v,\tau^A)+8\pi G~T(\eta,l)]~^2\epsilon,\label{omegaAchange}
\end{gather}
for arbitrary $\eta$, tangent to the cross-sections and lie dragged along $l$. At this point of time one might think that the change can be induced by two possible ways, 1) by assuming a time varying $\kappa$ 2) introduction of a non-zero stress energy tensor  or both 1) and 2). Note that both these terms are freely specifiable on $\mathcal H$, since there are no dynamical equations containing $D(\kappa)$ in the Newman Penrose equations or $D({\bf \Phi_{01}})$ in the conservations equations. In the foregoing discussing we will assume that $\kappa$ is a constant on $\mathcal H$. Since the area form is not changing with time one can equate the integrands on both sides of eq. (\ref{omegaAchange}), enabling the identification of a component of the stress energy tensor,
\begin{gather}
T(\partial_A,l)=\frac{1}{8\pi G}\partial_Af(v,\tau^B),
\end{gather}
where $f(v,\tau^B)$ is such that $\int_{v_1}^{v_2}f(v,\tau^A)=\mathcal F(\tau^A)$.
We further need to look for consistency conditions such that the conservation equation is obeyed by the stress tensor. If we get non zero values for any other component of the stress energy tensor then they must obey the conditions imposed on $\mathcal H$ through Einsteins' equations. For example $T(l,l)=0$ because the area is not changing. Consequently, the Dominant energy condition is definitely violated by the appearance of $T(\partial_A,l)$. One might think that by assuming $\kappa$ to be varying this pathology could have been avoided. It would be apparent that it is not the case, due to appearance of a $T(l,n)$ in the latter calculations. 
%
\subsection{Construction of a metric in the neighborhood of $\mathcal H$}
In this section we will try to reconstruct the metric in the neighborhood of $\mathcal H$. This will allow us to check, or impose further conditions on the stress energy tensor, so that the metric in the neighborhood of $\Delta_2$ is indeed that obtained by a supertranslation of the metric in the neighborhood of $\Delta_1$. We will make a number of gauge choices on $\mathcal H$ so as to make the calculations less cumbersome. In general the calculation can proceed without such gauge choices at the expense of changing the form of the final expression. In fact such a gauge choice can always be made in the neighbourhood of any null surface. The calculation will closely follow that of \cite{Krishnan:2012bt,Pawlowski:2003ys,Lewandowski:2014nta} with the relaxation of the weakly condition. In this section we will make use of the Newman Penrose formalism. Let us therefore recall the commutation relations between the null bases and spell out the gauge choices.
\begin{subequations}\label{comm}
	\begin{gather}
	(\Delta D - D\Delta)f = (\epsilon+\bar{\epsilon})\Delta f + (\gamma
	+ \bar\gamma)Df - (\bar\tau + \pi)\delta f -(\tau + \bar{\pi})\bar{\delta}f\,,\\
	(\delta D-D\delta)f =(\bar{\alpha}+\beta-\bar\pi)Df + \kappa\Delta
	f - (\bar{\rho}+\epsilon-\bar{\epsilon})\delta f
	-\sigma\bar{\delta}f\,,\\
	(\delta\Delta-\Delta\delta)f =-\bar\nu Df +
	(\tau - \bar{\alpha}-\beta)\Delta f + (\mu-\gamma+\bar\gamma)\delta f + \bar{\lambda}\bar{\delta}f\,,\\
	(\bar{\delta}\delta-\delta\bar{\delta})f = (\bar{\mu}-\mu)Df +
	(\bar{\rho}-\rho)\Delta f + (\alpha-\bar{\beta})\delta f -
	(\bar{\alpha}-\beta)\bar{\delta}f\,,
	\end{gather}
\end{subequations}
where $D=l^\mathfrak a\nabla_\mathfrak a, \Delta=n^\mathfrak a\nabla_\mathfrak a, \delta=m^\mathfrak a\nabla_\mathfrak a$. One can choose a coordinate $v$ on $\mathcal H$ such that $\lie_lv=1$. If we choose the function $f=v$, then from the above commutation relations it follows that $\mu=\bar\mu$ and $\pi=\alpha+\bar\beta$. Futher, one can choose $n$ to be an affinely parametrised geodesic and $m,\bar m, l$ to be parallely propagated along $n$ thus extending the null basis to a neighborhood of $\mathcal H$. On imposing such conditions it follows that \cite{Krishnan:2012bt}, $\gamma=\tau=\nu=0$.
The null netrads may then be expanded as,
\begin{subequations}
\begin{gather}
D=\nabla_l=\frac{\partial}{\partial v}+L(r,v,\tau^A)\frac{\partial}{\partial r}+T^A(v,\tau^A)\frac{\partial}{\partial \tau^A}\\
\delta=\nabla_m=M(v,\tau^A)\frac{\partial}{\partial r} +P^A(v,\tau^A)\frac{\partial}{\partial \tau^A}\\
\bar{\delta}=\nabla_{\bar m}=\bar{M}(v,\tau^A)\frac{\partial}{\partial r}+\bar{P}^A(v,\tau^A)\frac{\partial}{\partial \tau^A}
\end{gather}
\end{subequations}
 These coordinates are similar to the Bondi coordinates
	at null infinity except that there is no choice of a preferred Bondi frame. Hence $n$ may have non zero shear. In this coordinate system $\mathcal H$ is assumed to be foliated by the level surfaces of $v$, which may be denoted as $S_{\mathcal H_v}$. At each $S_{\mathcal H_v}$ the past out-going light cone is generated by $n$ and is parametrised by $r$. Thus we have, $\underline n=-dv$ and the directional derivative along $n$, $\Delta=n^{\mathfrak a}\nabla_{\mathfrak a}=-\frac{\partial}{\partial r}$. Denoting the nth order (in $r$) of a Newman Penrose scalar as ~$\accentset{(n)}{}$ ~, the boundary conditions on the $\mathcal H$ reads,
	\begin{equation}
	\accentset{(0)}{\rho}=0,\;\;\accentset{(0)}\kappa=0.
	\end{equation}
	The spin coefficient ~$\accentset{(0)}{\epsilon}+\accentset{(0)}{\bar{\epsilon}}$ ~will be chosen to be a constant thus implying that the surface gravity~ $\kappa=\accentset{(0)}{\epsilon}+\accentset{(0)}{\bar{\epsilon}}$ ~is a constant on $\mathcal H$. The functions $L$, $T^A$, $M$, and $P^A$ are determined from the commutation relations eq. (\ref{comm}), by substituting in turn, $r$ and $\tau^A$ and integrating the radial differential equations thus obtained. The following expansions are obtained by following such a procedure.
	%
\begin{subequations}
\begin{gather}
L=\kappa_{H}~r+\mathcal O(r^2)\\
M=\accentset{(0)}{\bar{\pi}}(v,\tau^A)~r+\mathcal O(r^2)\\
T^A=r~\bigg[~\accentset{(0)}\pi(v,\tau^A)\accentset{(0)}P^A+\bar{\pi}^0(v,\tau^A)\accentset{(0)}{\bar{P}}^A\bigg]+\mathcal O(r^2)\\
P^A=\accentset{(0)}{P}^A+r~\bigg[\accentset{(0)}{\mu}(v,\tau^A)~\accentset{(0)}{P}^A+\accentset{(0)}{\bar{\lambda}}(v,\tau^A)~\accentset{(0)}{\bar{P}}^A\bigg]+\mathcal O(r^2),
\end{gather}
\end{subequations}
The	null co-tetrads can easily be obtained from the above expansions, allowing us to write the metric to linear order in $r$. It has the form,
	\begin{equation}
	ds^2=2dvdr-2r\big(\omega_A(v,\tau^A)dvd\tau^A+{\kappa}_Hdv^2\big)+\big(\accentset{(0)}{q}_{AB}(\theta,\phi)+\accentset{(1)}{q}_{AB}(v,\theta,\phi)r\big)d\tau^Ad\tau^B+\mathcal O(r^2)
	\end{equation}
	where $\omega_A(v,\tau^A)=\big(\pi^0(v,\tau^A)P_A^0+\bar{\pi}^0(v,\tau^A)\bar{P}_A^0\big)$.
	Note that $\omega_A$ in general is of the form $\omega_A^{\Delta_1}+\kappa~\partial_A\int^v f~dv:=\omega_A^{\Delta_1}+\mathcal \partial_AG(v,\tau^A)$, where the superscript in $\omega_A^{\Delta_1}$ denotes the value at the initial WIH. For the case where $\Delta_1$ is of Type I or spherically symmetric, $\omega_A^{\Delta_1}$ is trivial. Hence in such a case we have, $\omega_A=\partial_A G$. The next to leading order metric $\accentset{(1)}{q}_{AB}$ can be obtained by using the expression for the extrinsic curvature $K^{(n)}_{AB}$. Conversely, it can also be obtained by solving for $\mu,~\lambda$. Since we are interested in the coordinate components we will instead use $K^{(n)}_{AB}$. In particular we require it to be of the form obtained in eq.(\ref{WTEC}),
	so that the final metric matches with that of the supertranslated one. So, let us start by trying to solve the first order differential equation determing $K^{(n)}_{AB}$ eq. (\ref{ETEC}). The solution of the differential equation can be written as,
	\begin{gather}
	K^{(n)}_{AB}=\frac{1}{\kappa}\bigg[\frac{1}{2}~^2\mathcal R(\partial_A,\partial_B)-\mathcal D_{(A}\omega^{\Delta_1}_{B)}-\omega^{\Delta_1}_{(A}\omega^{\Delta_1}_{B)}\bigg]_{v_1}^{v}+e^{-\kappa (v-v_1)}\int_{v_1}^{v}\bigg[\frac{1}{2}R(\partial_A,\partial_B)-P_{AB}\bigg]e^{\kappa v}dv,\label{Q1AB}
	\end{gather}
	where $P_{AB}(v):=\mathcal D_{(A}\mathcal D_{B)}G+\mathcal D_{(A}G~\omega^{\Delta_1}_{B)}+\omega^{\Delta_1}_{(A}\mathcal D_{B)}G+\mathcal D_{(A}G~\mathcal D_{B)}G$. An integration by parts of the second term, gives,
	\begin{gather}
\int_{v_1}^{v}\bigg[\frac{1}{2}R(\partial_A,\partial_B)-P_{AB}\bigg]e^{\kappa v}dv=-\bigg[\frac{e^{\kappa v}}{\kappa}P_{AB}\bigg]_{v_1}^v+\int_{v_1}^v
\frac{e^{\kappa v} }{\kappa}\bigg(\frac{\kappa}{2}R(\partial_A,\partial_B)+\dot P_{AB}\bigg)
dv\label{RAB}
	\end{gather}
The first three terms in eq. (\ref {Q1AB}) arise due to normal evolution without the inclusion of the region $\mathcal H$. One now has to choose $R(\partial_A,\partial_B)$ in such a way that the next to leading order metric at the final slice $v_2$ matches with the supertranslated one. One of the choices is to set the term under the integral in eq. (\ref{RAB}) equal to zero i.e,
\begin{gather}
R(\partial_A,\partial_B)=-2\bigg[\mathcal D_{(A}\mathcal D_{B)}f+\mathcal D_{(A}f~\omega^{\Delta_1}_{B)}+\omega^{\Delta_1}_{(A}\mathcal D_{B)}f+\kappa\mathcal D_{(A}f~\mathcal D_{B)}f\bigg]\label{RAB2}
\end{gather}
A contraction of eq. (\ref{RAB2}) with $q^{AB}$ then yields the following expression for a component of the stress energy tensor,
\begin{gather}
T(l,n)\stackrel{\mathcal H}{=}-\frac{1}{8\pi G}\bigg[\mathcal D^2f+q^{AB}\omega^{\Delta_1}_B\mathcal D_Af+\kappa||\mathcal Df||^2\bigg]
\end{gather} 
We are now in a position to discuss the case with varying $\kappa$. If we chose a non-constant $\kappa$, clearly the first three terms in eq. (\ref{Q1AB}) would not have been reproduced. Further, a contribution to $T(l,n)$ would have arised even in such a case, leading to a violation of the dominant energy for negative values of $T(l,n)$.
%
\subsection{Analysis of Conservation equation in Newman Penrose basis}\label{convpen}
In this section we will analyse the conservation equation. This will allow us to partially reconstruct the stress energy tensor in the neighborhood of $\mathcal H$. In a Newman Penrose basis the conservation equation can be written as (appendix \ref{CONSNP}),
\begin{gather}
-\lie_{l}T(n,Z)-\lie_{n}T(l,Z)+T(\nabla_nl,Z)+T(\nabla_ln,Z)+q^{AB}\nabla_{\partial_A}T(\partial_B,Z)-q^{AB}T(\nabla_{\partial_A}\partial_B,Z)\nn
+T(n,\nabla_lZ)+T(l,\nabla_nZ)-q^{AB}T(\partial_A,\nabla_{\partial_B} Z)=0
\end{gather}
In the above expression we will choose $Z$ to be different vector fields and try to analyze the equations so that the stress energy tensor could be consistently found. Let us choose $Z=l$, the vector field generating $\mathcal H$, and evaluate the expression on $\mathcal H$. In such a case one has,
\begin{gather}
%
%
-\lie_{l}T(n,l)-\lie_{n}T(l,l)+\mathcal D^A\bigg(T(\partial_A,l)\bigg)=0,
\end{gather}
where we have used the boundary conditions axund the gauge choices. On replacing the expressions previously obtained one gets,
\begin{gather}
T(l,l)=\frac{r~\kappa}{8\pi G}\bigg[\mathcal D^2\dot f+\omega_A~q^{AB}\mathcal D_B\dot f+2~D^A\dot f~\mathcal D_A f+\mathcal D^2f\bigg],
\end{gather}
where $\dot f$ implies a derivative w.r.t $v$. In the case of spherical symmetry, we have $\omega_A^{\Delta_1}=0$. Further, if we want only linearised supertranslations to be induced then the term quadratic in $f$ can be ignored. If the process is a through a shock wave, then $f$ is separable into a delta function and a functions on the sphere which can be expanded in spherical harmonics. The above equation then gives,
\begin{gather}
T(l,l)=\frac{r~\kappa}{4\pi G}\bigg[l(l+1)Y_{lm}\delta(v-v_0)\bigg]+\mathcal O(r^2),
\end{gather}
where we have ignored the derivative of the delta function, as it does not contribute to the integrated change. These are the only components that can be determined from the first order expansion. Determining any other component requires the expansion of the metric upto second order, which we will address in a future work. Here, we will only write down the other conservation equations and discuss the consequences. When, $Z=n$, we have,
\begin{gather}
-\lie_l T(n,n)-\lie_{n}T(l,n)+2\kappa ~T(n,n)+2\omega_Aq^{AB}T(\partial_B,n)\nn
+\mathcal D^AT(\partial_A,n)-\omega_Bq^{AB}T(\partial_A,~n)-T(\partial_A,W_n(\partial_B))q^{AB}=0
\end{gather}
The components of the stress energy tensor arising in the above equation can all be non zero on an isolated horizon. In particular the corresponding Einstein's equation contain the components of the metric at second order in expansion around $\Delta$. Hence these can only be determined if specific conditions are imposed at second order in the expansion \cite{Lewandowski:2000nh,Lewandowski:2014nta}. This is also the case for the metric in the neighborhood of $\mathcal H$. The components that have already been determined, however, do not depend on such choices. Hence these equation do not affect the components already found. Let us assume that $T(n,n),~T(\partial_A,n)$ and $T(\partial_A,\partial_B)$ are trivial. This implies the the radial derivative of $T(l,n)$ is zero. Thus determining $T(l,n)$ to next order. Finally a contraction with $\partial_C$ gives,
\begin{gather}
%
%
-\lie_lT(n,\partial_C)-\lie_{n}T(l,\partial_C)-\kappa~ T (n,\partial_C)+\omega_Aq^{AB}T\big(\partial_B,\partial_C\big)-T(l,W_{n}(\partial_C))+\mathcal D^AT_{AC}=0
\end{gather}
If the above conditions are also imposed in this equation, the component $T(l,\partial_A)$ can be computed to next order in the expansion in the neighborhood of $\mathcal H$. In the following section we will further analyse the boundary conditions in Newman Penrose basis and look for inconsistencies if any.
\subsection{Analysis of boundary conditions}
To analyse the geometry of $\mathcal H$, let us calculate the exterior derivative of $\omega$. Since this underges a nontrivial change under the action of the supertranslations it is useful to obtain an expression for $d\omega$. In the case of an isolated horizon $d\omega_{IH}=(\Im{\bf\Psi_2})~^2\epsilon$. However unlike an isolated horizon the induced connection $(\omega)$ cannot be Lie-dragged along $\mathcal H$.  We start with the definition of Riemann tensor $(\nabla_{\partial_a}\nabla_{\partial_b}-\nabla_{\partial_b}\nabla_{\partial_a})X=R(\partial_a,\partial_b)X$ as in \cite{Ashtekar:2001is}. Let us choose  $X=l$. Now, consider the left hand side of the above equation. Using the expression for $\nabla_{\partial_a}l$ given in appendix (\ref{NP}) we get,
\beq
\nabla_{\partial_a}(\nabla_{\partial_b}l)-\nabla_{\partial_b}(\nabla_{\partial_a}l)&=&\nabla_{\partial_a}\bigg(\omega\big(\partial_b\big)~l-(\gamma+\bar{\gamma})~\underline l\big(\partial_b\big)~ l+\bar \tau~ \underline{l}\big(\partial_b\big)~m+ \tau~ \underline{l}\big(\partial_b\big)~\bar m\bigg)-(a\leftrightarrow b)\nn
\eeq
 Contracting the above by $n$ and using the gauge choices, gives,
\beq
g(n,R(\partial_a,\partial_b)~l)&\stackrel{\Delta}{=}-2d\omega
\eeq
Now the Weyl tensor can be written in terms of the curvatures as in appendix eq. (\ref{Weyl}). Expanding each of the terms in a Newman Penrose basis, one gets the following expression,
\beq
d\omega&=&(\Im {\bf\Psi_2})~^2\epsilon+{\bf\Phi_{01}}~n\wedge \bar m+ {\bf\bar\Phi_{10}}~n\wedge m
\eeq
If one now assumes that the dominant energy condition holds that is ${\bf\Phi_{01}}\stackrel{\Delta}{=}0$. Then the weakly condition implies that the surface gravity $\epsilon+\bar{\epsilon}=l.\omega$ is a constant on the horizon. However if one weakens the conditions by allowing a violation of the dominant energy condition, then the weakly condition and the zeroth law are independent. In fact in the case of $\mathcal H$ we have assumed that the zeroth law continues to hold (if the constancy of $\kappa$ can be interpreted as a zeroth law) but the dominant energy condition is violated. This implies that the pull back of $\omega$ on to the leaves of $\mathcal H$, $S_{\mathcal H}$, can change with time.

We will now be interested in some of the Newman Penrose equations. Not all are important in this context as they might contain radial derivatives of the connection coefficients. The important ones are mostly those which contain derivatives tangential to $\mathcal H$. Now recall the following Newman Penrose equation,
\begin{gather}
\delta\rho-\bar{\delta}\sigma=\rho(\bar{\alpha}+\beta)-\sigma(3\alpha-\bar{\beta})+\tau(\rho-\bar{\rho})+\kappa(\mu-\bar{\mu})-{\bf\Psi_1}+{\bf\Phi_{01}}
\end{gather}
For the boundary conditions to hold on $S_{\mathcal H}$ one must therefore have a non-zero $\bf\Psi_1$ such that $\bf\Psi_1={\bf\Phi_{01}}$. One of the important consequences of the boundary condition, which is an evolving rotation one form, is captured in the following equation,
\begin{gather}
D(\alpha+\bar\beta)=(\alpha+\bar\beta)(\epsilon-\bar\epsilon)+2\Re{\bf\Phi_{01}}
\end{gather} 
This is a combination of two Newman Penrose equations. One can make a rotation in the $m,~\bar m$ frame to make $\epsilon=\bar{\epsilon}$ on $\mathcal H$. Thus, the above equation describes the evolution of the rotation one form and is the main indicator of an evolving system. Other equations either trivially hold or put conditions on the radial derivative of a Newman Penrose scalar. Let us now go over to the Bianchi identities. We will only consider those which have a potential to give rise to discrepancies due to the assumptions made. Consider a contracted Bianchi identity,
\begin{gather}
\bar{\delta}{\bf \Phi_{01}}+\delta{\bf \Phi_{10}}-D({\bf \Phi_{11}}+3{\bf \Lambda})-\triangle{\bf \Phi_{00}}=\bar{\kappa}{\bf \Phi_{12}}+\kappa {\bf \Phi_{21}}+(2\alpha+2\tau-\pi){\bf \Phi_{01}}\nn
+(2\bar\alpha+2\bar\tau-\bar\pi){\bf \Phi_{10}}-2(\rho+\bar{\rho}){\bf \Phi_{11}}-\bar\sigma
{\bf \Phi_{02}}-\sigma{\bf \Phi_{20}}+[\bar{\mu}+\mu-2(\gamma+\bar{\gamma})]{\bf \Phi_{00}}
\end{gather}
On putting the boundary conditions it follows that,
\begin{gather}
\bar{\delta}{\bf \Phi_{01}}+\delta{\bf \Phi_{10}}-D({\bf \Phi_{11}}+3{\bf \Lambda})-\triangle{\bf \Phi_{00}}=-(\pi-2\alpha){\bf \Phi_{01}}-(\bar\pi-2\bar{\alpha}){\bf \Phi_{10}}
\end{gather}
This is nothing but one of the stress energy conservation equations and relates the radial derivative of $T(l,l)$ with the value of $T(l,n)$ and $T(\partial_A,l)$ on $\mathcal H$. In section (\ref{convpen}) it is essentially this equation that we have solved in order to find the value of $T(l,l)$ in the neighborhood of $\mathcal H$. Let us now consider the following Bianchi identity,
\begin{gather}
-\bar \delta {\bf \Psi_0}-D{\bf \Psi_1}+(4\pi-\alpha){\bf \Psi_0}-2(2\rho+\epsilon){\bf \Psi_1}+3\kappa{\bf \Psi_2}-D{\bf \Phi_{01}}+\delta {\bf \Phi_{00}}\nn
+2(\epsilon+\bar{\rho}){\bf \Phi_{01}}+2\sigma {\bf \Phi_{10}}-2\kappa {\bf \Phi_{11}}-\bar{\kappa}{\bf \Phi_{02}}+(\bar{\pi}-2\bar{\alpha}-2\beta){\bf \Phi_{00}}=0
\end{gather}
On using the condition ${\bf \Psi_1}\stackrel{\mathcal H}{=}{\bf \Phi_{01}}$, and ${\bf \Phi_{00}}\stackrel{\mathcal H}{=}0$, this is clearly seen to be satisfied. Next consider the equation.
\begin{gather}
\bar \delta {\bf \Psi_1}-D{\bf \Psi_2}-\lambda{\bf \Psi_0}+2(\pi-\alpha){\bf \Psi_1}+3\rho{\bf \Psi_2}-2\kappa_{NP}{\bf \Psi_3}+\bar{\delta}{\bf \Phi_{01}}-\triangle {\bf \Phi_{00}}\nn
-2(\alpha+\bar{\tau}){\bf \Phi_{01}}+2\rho {\bf \Phi_{11}}+\bar{\sigma} {\bf \Phi_{02}}-(\bar \mu-2\gamma-2\bar\gamma){\bf \Phi_{00}}-2\tau {\bf \Phi_{01}}-2D{\bf \Lambda}=0
\end{gather}
This is again one of the Bianchi identitites. The assumed boundary conditions then imply that,
\begin{gather}
2\bar \delta {\bf \Phi_{01}}-2\delta {\bf \bar\Phi_{01}}+2(\pi-2\alpha){\bf \Phi_{01}}-2(\bar\pi-2\bar\alpha){\bf \bar\Phi_{01}}-D \Im{\bf \Psi_2}=0
\end{gather}
On using the expression for ${\bf \Phi_{01}}=\delta \mathcal F$ and the commutation relations, it follows that, $D \Im{\bf \Psi_2}=0$, which is again consistent with the fact that the $\Im{\bf \Psi_2}$ should be preserved in time so that the amgular momentum multipole moments do not change. Let us now take the real combination,
\begin{gather}
2\bar \delta {\bf \Phi_{01}}+2\delta {\bf \bar\Phi_{01}}+2(\pi-2\alpha){\bf \Phi_{01}}+2(\bar\pi-2\bar\alpha){\bf \bar\Phi_{01}}-\triangle{\bf \Phi_{00}}-2D\Re{\bf \Psi_2}-2D{\bf \Lambda}=0
\end{gather}
The conservation equation then implies that, $D\bigg(\Re{\bf \Psi_2}+\frac{R(l,n)}{2}-\frac{~^2\mathcal R}{4}\bigg)=0$. Now recall that the boundary conditions imply that $D~^2\mathcal R=0$. The equation then takes the form,
\begin{gather}
D\bigg(\Re{\bf \Psi_2}+8\pi G~\mathcal T\bigg)=0,
\end{gather}
where $\mathcal T=q^{AB}T(\partial_A,\partial_B)$. In the absence of the components $T(\partial_A,\partial_B)$ this reduces to $D(\Re{\bf\Psi_2})=0$, which is again consistent with our assumptions.
\section{Conclusions}
In the context of asymptotically flat spacetime at null-infinity the symmetry group is the BMS group. It is known that the spacetimes related by BMS transformations represent inequivalent vacua of the gravitational field. One may therefore ask the question what processes induce a transition between two such spacetimes. Interestingly black hole horizons also admit an infinite dimensional symmetry group. Due to the null nature of a black horizon this is similar to the BMS group. However it is a matter of debate whether it should be the full BMS group or only a sub-group of it, and depends on the kind of boundary condition one chooses. The emergence of supertranslation symmetry on the horizon is however obtained even with the strictest of boundary conditions and beyond any debate. In this paper we revisited such a symmetry analysis in the context of isolated horizons. Though this has been discussed previously, our approach here makes use of the boundary data only and avoids making use of the metric in the neighborhood.

We then go on to find a process that induces a supertranslation transition in analogy with transitions between spacetimes related by a BMS transformation \cite{Compere:2018ylh}. Transformation between black holes related by supertranslations have been studied before \cite{Hawking:2016sgy}. However these are asymptotic supertranslations acting on black hole spacetimes. In our case the supertranslations are those obtained by preserving certain boundary conditions on the horizon. As has been pointed out before there is no reason to believe that the extensions of horizon supertranslation generators into bulk will necessarily reproduce the supertranslations at $\mathscr I^-$. Therefore the content of the stress energy tensor obtained here may not match with that obtained in \cite{Hawking:2016sgy}. Certain essential features are however seen to be reproduced e.g it contains first and second derivatives of the function generating the supertranslation and the stress energy tensor is seen to violate the dominant energy condition on the horizon.

It is imperative that we discuss the pathological nature of the stress energy tensor. The dominant energy condition is an essential input in the proof of the Black hole topology theorem \cite{Hawking:1971vc,Galloway:2005mf}. At this point of time it is difficult to comment on the fate of the Topology Theorem if the phase space of solutions is allowed to contain black holes related by supertranslations. There is however no reason to believe that such supetranslation transitions can be induced by classical matter sources. An evaporating black hole e.g must have a negative energy flux across the horizon. While an exact solution, of Einsteins' equations, describing such a process can be obtained, the stress energy tensor necessarily violates the null energy condition. Such classically pathological stress energy tensor is however known to arise is a semi-classical framework of quantum fields on curved spacetimes. The regularised stress energy tensor is known to violate the null energy condition thus accounting for the Hawking flux and black hole evaporation. The violation of the dominant energy condition is also a plausible outcome. In such a case the transition would be driven by semi-classical effects. The result obtained in \cite{Chu:2018tzu}, where it is shown that the spectrum of Hawking radiation from a supertranslation of Vaidya spacetime does carry information about the supertranslation hair, further lends support to this argument.

As an immediate corollary of the above fact we can argue that if two supertranslated weakly isolated horizons are taken to be in the same phase space then it necassary to include a Dominant energy violating flux. Alternatively if they are taken to be in different disconnected components of the phase space then it is impossible to distinguish them by some classical measurement only, which does allows the measurement of super-rotation charges only upto a supertranslation. Hence a measuring scheme that allows the measurement of the Dominant energy condition violating flux is required to distinguish between such supertranslated weakly isolated horizons.  

Finally, we would like to point out the drawbacks and improvements of our method. As has been pointed out before we only try to identify a stress energy tensor that induces a transition between supertranslated black holes avoiding  general dynamical evolution where there might be other nontrivial fluxes. In general the part of the stress energy tensor inducing only a supertranslation, the analogue of the soft flux at null infinity, may be mixed with other non-trivial fluxes, making it diffcult to identify it. Our approach therefore has been minimalistic. In a future work we hope to address the problem of general dynamical evolutions in the Dynamical Horizon framework.

In the current context one might be able to discuss the transition completely on the phase space of non-expanding horizon that includes radiative solutions as in \cite{Sousa:2017auc}. The phase space of boundary data on a non-expanding horizon (in this case $\Delta_1\cup\mathcal H\cup\Delta_2$) therefore needs to be constructed with the inclusion of matter as opposed to gravitational data alone. This might allow a precise formulation of such aupertranslation transformations.
\section*{Acknowledgements}
A.G acknowledges Srijit Bhattacharjee and Prof. Li-Ming Cao for fruitful discussions. A.G also acknowledges the support by a grant from the NSF of China with Grant No: 11947301 and the hospitality at Saha Institute of Nuclear Physics, where part of this work was done.
\appendix
\section{Notations and conventions}\label{N&C}
 The covariant derivative $\nabla: T\mathcal M \otimes T\mathcal M\rightarrow T\mathcal M$ for two vector fields $W,Z~\in ~T\mathcal M$ will be denoted by $\nabla_WZ$. Let $\mathcal S$ be an immersed submanifold. The tangent space at the point $x\in\mathcal S$ can be decomposed as $T_x\mathcal M=T_x\mathcal S\oplus T_x^{\perp}\mathcal S$. The covariant derivative on $\mathcal S$ will be denoted by $D_XY$ for $X,Y~\in T\mathcal S$. Gauss decomposition then allows us to write,
\begin{gather}
\nabla_XY=D_XY+K(X,Y),
\end{gather}
where $K(X,Y)$ is the extrinsic curvature. We denote as $\nabla^{\perp}_XN^\perp$, the connection in the  normal bundle, where $X\in T\mathcal S$ and $N^\perp\in T^\perp\mathcal S $. The definition of the shape operator $W_{N^\perp}(X)$ follows,
\begin{gather}
\nabla_XN^\perp=\nabla_X^\perp N^\perp-W_{N^\perp}(X).
\end{gather} 
A relation between the shape operator and the extrinsic curvature follows,
\begin{gather}
g(W_{N^\perp}(X),Y)=g(N^\perp,K(X,Y)),
\end{gather}
where $X,Y\in T\mathcal S$ and $N^\perp\in T^\perp\mathcal S$. The Riemann tensor is defined as,
\begin{gather}
R(W,U)V\equiv[\nabla_W,\nabla_U]V-\nabla_{[W,U]}V
\end{gather}
Similarly one can define an intrinsic Riemann tensor as,
\begin{gather}
\mathcal R(X,Y)Z\equiv[D_X,D_Y]Z-D_{[X,Y]}Z
\end{gather}
We write down the equations of Gauss and Codazzi, in this notation. Let $X,Y,Z,W\in T\mathcal S$ and $N^\perp\in T^\perp\mathcal S$. Then the Gauss equation is given as,
\begin{gather}
g(R(X,Y)Z,W)=g(\mathcal R(X,Y)Z,W)-g(K(X,Z),K(Y,W))+g(K(X,W),K(Y,Z)),\label{Gauss}
\end{gather}
and the Codazzi equation as,
\begin{gather}
g(R(X,Y)N^\perp,Z)=g((\nabla_YK)(X,Z),N^\perp)-g((\nabla_XK)(Y,Z),N^\perp)\label{Codazzi}
\end{gather}
\section{The Connection in terms of Newman-Penrose co-effecients}\label{NP}
	In the Newman Penrose formalism one chooses all four tetrads to be null 
	vectors $(l,n,m,\bar{m})$, $l$ and $n$ being real null vector, whereas $m$ is a complex null vector 
	$\bar{m}$ being its complex conjugate. For a $(-+++)$ signature these null tetrads have to satisfy following conditions.
	\begin{equation}
	l.n=-1\;\; \mathsf{and}\;\; m.\bar{m}=1
	\end{equation}
	\begin{equation}
	l.l=n.n=m.m=m.\bar{m}=0
	\end{equation}
	The space time metric is given by 
	\begin{equation}
	g(\partial_{\mathfrak{a}},\partial_{\mathfrak b})=-\underline {l}\big(\partial_{(\mathfrak{a}}\big)\underline {n}\big(\partial_{\mathfrak b)}\big)+\underline{m}\big(\partial_{(\mathfrak {a}})\underline{\bar{m}}\big(\partial_{\mathfrak b)}\big)
	\end{equation}
	Directional derivative along basis vectors are
	
	\begin{equation}
	D=l^{\mathfrak a}\nabla_{\partial_{\mathfrak a}},\;\;\;\Delta=n^{\mathfrak a}\nabla_{\partial_{\mathfrak a}},\;\;\;\delta=m^{\mathfrak a}\nabla_{\partial_{\mathfrak a}},\;\;\;\;\bar{\delta}=\bar{m}^{\mathfrak a}\nabla_{\partial_{\mathfrak a}} 
	\end{equation}
	Directional derivative of the tetrads are.
	\begin{subequations}
		\begin{gather}
			Dl=(\epsilon+\bar{\epsilon})l-\bar{\kappa}m-\kappa\bar{m},~~	Dn=-(\epsilon+\bar{\epsilon})n-\pi m-\bar{\pi}m,~~Dm=\bar{\pi}l-\kappa n-(\epsilon-\bar{\epsilon})m\\
			\Delta l=(\gamma+\bar{\gamma})l-\bar{\tau}m-\tau\bar{m},~~\Delta n=-(\gamma+\bar{\gamma})n-\nu m-\bar{\nu}\bar{m},~~\Delta m=\bar{\nu}l-\tau n+(\gamma-\bar{\gamma})m\\
			\delta l=(\bar{\alpha}+\beta)l-\bar{\rho}m-\sigma\bar{m},~~\delta n=-(\bar{\alpha}+\beta)n-\mu m-\bar{\lambda}\bar{m},~~			\delta m=\bar{\lambda}l-\sigma n -(\beta-\bar{\alpha})m\\
			\bar{\delta} m=\bar{\mu}l-\rho n-(\alpha-\bar{\beta})m
		\end{gather}
	\end{subequations}
To obtain the connection in terms of the Newman-Penrose coeffecients one fixes a set of internal null vectors $(l_I,n_I,m_I,\bar m_I) $ on $\Delta$ such  
that $\partial_a (l_I,n_I,m_I,\bar m_I)\stackrel{\Delta}{=}0$. For any co-tetrad 
$\underline{e^I}$, the null co-tetrad $(\underline l,~\underline n_,~\underline m_,~\underline{\bar m})$ can then be expanded as 
$\underline l=\underline {e^I}~l_I$, thus providing an expression for $\Sigma_{IJ}$.
\beq
\underleftarrow{\Sigma}^{IJ}\stackrel{\Delta}{=}2l^{[I}n^{J]}~^{2}\epsilon+2n\wedge(im~l^{[I}\bar
{m}^{J]}-i\bar{m}~l^{[I}m^{J]})
\eeq 
The connection can be written in terms of the Newman- Penrose
coefficients.
\begin{gather}
A_{IJ}\stackrel{\Delta}{=}2\left[(\epsilon+\bar{\epsilon})\underline n+(\gamma+\bar{
	\gamma})\underline l-(\bar{\alpha}+\beta)\underline {\bar{m}}-(\alpha+\bar{\beta} )\underline m\right]\, 
l_{[I}n_{J]}\nn
+2\left[-\bar{\kappa}\underline n-\bar{\tau}\underline l+\bar{\rho}\underline{\bar{m}}+\bar{\sigma}
\underline m\right]\, m_{[I}n_{J]}+2\left[-{\kappa}\underline n-\tau 
\underline l+{\rho}{\underline m}+{\sigma}\underline{\bar{m}}\right]\, \bar{m}_{[I}n_{J]}\nn
+2\left[\pi \underline n+\nu \underline l-\mu\underline{\bar{m}}-\lambda \underline m\right]\, 
m_{[I}l_{J]}+2\left[\bar{\pi} 
\underline n+\bar{\nu}\underline l-\bar{\mu}{\underline m}-\bar{\lambda}\underline{\bar{m}}\right]\, 
\bar{m}_{[I}l_{J]}\nn
+2\left[-(\epsilon-\bar{\epsilon})\underline n-(\gamma-\bar{\gamma})\underline l+({\alpha}-\bar\beta){\underline m}+(\beta-\bar{
	\alpha})\underline{\bar{m}}\right]\, m_{[I}\bar{m}_{J]}\label{connection}
\end{gather} 
\section{Conservation equation in Newman Penrose basis}\label{CONSNP}
In this section we would like to write the conservation equation in a Newman Penrose basis. This will be used in the main text to analyse the content of the stress energy tensor. First, note that,
\begin{gather}
\nabla_{X}T(Y,Z)=(\nabla_XT)(Y,Z)+T(\nabla_XY,Z)+T(X,\nabla_XZ)
\end{gather}
Now let $X=\partial_\mathfrak a$ and $Y=\partial_\mathfrak b$. Contracting both side by $g^{\mathfrak a\mathfrak b}$ we have, for the left hand side
\begin{gather}
\Bigg(-l^\mathfrak a n^\mathfrak b-l^\mathfrak b n^\mathfrak a+q^{AB}\bigg(\frac{\partial}{\partial{\tau^A}}\bigg)^\mathfrak a\bigg(\frac{\partial}{\partial{\tau^B}}\bigg)^\mathfrak b\Bigg)\nabla_{\partial_\mathfrak a}T(\partial_\mathfrak b,Z)
\end{gather}
Consider the first, second and the third term in the above expression can be written as,,
\begin{gather}
\bigg(-l^\mathfrak a n^\mathfrak b\bigg)\nabla_{\partial_\mu}T(\partial_\nu,Z)=-\nabla_lT(n,Z)+\nabla_ln^\mathfrak b~T(\partial_\mathfrak b,Z)\\
\bigg(-n^\mathfrak a l^\mathfrak b\bigg)\nabla_{\partial_\mathfrak a}T(\partial_\mathfrak b,Z)=-\nabla_nT(l,Z)+\nabla_nl^{\mathfrak b}~T(\mathfrak b,Z)\\
\Bigg(q^{AB}\bigg(\frac{\partial}{\partial{\tau^B}}\bigg)^\mathfrak b\Bigg)\nabla_{\partial_{A}}T(\partial_\mathfrak b,Z)=q^{AB}\nabla_{\partial_A}T(\partial_B,Z)-q^{AB} T(\nabla_{\partial_A}\partial_B,Z)
\end{gather}
Now consider the terms on the right hand side. Note that, 
$g^{\mathfrak a\mathfrak b}(\nabla_{\partial_\mathfrak a}T)(\partial_\mathfrak b,Z)$ is zero by the conservation law, while the second and the third term gives,
\begin{gather}
\Bigg(-l^\mathfrak a n^\mathfrak b-l^\mathfrak b n^\mathfrak a+q^{AB}\bigg(\frac{\partial}{\partial{\tau^A}}\bigg)\Bigg)T(\nabla_{\partial_\mathfrak a}\partial_\mathfrak b,Z)\nn
=-n^\mathfrak b T(\nabla_l\partial_\mathfrak a,Z)-l^\mathfrak a T(\nabla_n\partial_\mathfrak a,Z)+q^{AB}\bigg(\frac{\partial}{\partial{\tau^B}}\bigg)^\mathfrak b T(\nabla_{\partial_A}\partial_\mathfrak b,Z)\\
\Bigg(-l^\mathfrak a n^\mathfrak b-l^\mathfrak b n^\mathfrak a+q^{AB}\bigg(\frac{\partial}{\partial{\tau^A}}\bigg)\Bigg)T(\partial_\mathfrak b,\nabla_{\partial_\mathfrak a}Z)=-T(n,\nabla_lZ)-T(l,\nabla_nZ)+q^{AB}T(\partial_A,\nabla_{\partial_B} Z)
\end{gather}
If all the terms are taken to the left hand side, we have the following equation,
\begin{gather}
-\lie_{l}T(n,Z)-\lie_{n}T(l,Z)+T(\nabla_nl,Z)+T(\nabla_ln,Z)+q^{AB}\nabla_{\partial_A}T(\partial_B,Z)-q^{AB}T(\nabla_{\partial_A}\partial_B,Z)\nn
+T(n,\nabla_lZ)+T(l,\nabla_nZ)-q^{AB}T(\partial_A,\nabla_{\partial_B} Z)=0
\end{gather}
This will be used in the main text for various choices of $Z$ and the resulting equation would be analysed for consistency and reconsctruction of the stress energy tensor in the neighborhood of $\mathcal H$.
%
%
\section{Evolution of the rotation one form}\label{ERO}
In this section we will re-derive the evolution of the rotation one -form in the notation used in this manuscript. The derivation genralises the results obtained in \cite{Booth:2006bn} by the inclusion of vector fields tangent to a marginally trapped surface. In the main text we will however not be requiring this generalisation. Let us first recall its definition, $\omega_A:=-g(n,\nabla_{\partial_A}l)$. Since $\omega_A$ is a number in spacetime (it is however a one form under coordinate transformation of $S_{\mathcal H}$) taking its Lie derivative is same as taking its covariant derivative,
\begin{gather}
-\nabla_{X}\omega_A
=g(\nabla_Xn,\nabla_{\partial_A}l)+g(n,R(X,\partial_A)l~)+\nabla_{\partial_A}g(n,\nabla_X l)
-g(\nabla_{\partial_A}n,\nabla_X l)\nn
\nn
=K^{(n)}(X^T,\partial_C) q^{CD}K^{(l)}_{DA}-K^{(l)}(X^T,\partial_C) q^{CD}K^{(n)}_{DA}-(\partial_C\alpha+\alpha\omega_C) q^{CD}K^{(l)}_{DA}\nn
-(\partial_C\beta+\beta\omega_C) q^{CD}K^{(n)}_{DA}+g(n,R(X,\partial_A)l~)-\partial_A\kappa_X,
\end{gather}
where we have expanded the normal part of the evolution vector as $\alpha~l-\beta~n$.
Let us now try to write down the Riemann tensor in terms of quantities realizable through Einsteins' equations. 
Contracting eq.~(\ref{Codazzi}) with $q^{CB}$ we have,
\begin{gather}
g(l,R(\partial_A,l)n)-R(l,\partial_A)=\mathcal D^BK^{(l)}_{BC}-\mathcal D _AK^{(l)}+\omega ^BK^{(l)}_{BC}-\omega _AK^{(l)}\nn
g(n,R(\partial_A,n)l)-R(n,\partial_A)=\mathcal D ^BK^{(n)}_{BC}-\mathcal D _AK^{(n)}+\omega ^BK^{(n)}_{BC}-\omega _AK^{(n)}\label{codazzi3}
\end{gather}
Thus we can write 
\begin{gather}
g(n,R(X,\partial_A)l~)=-R(X,\partial_A)+\alpha\bigg(\mathcal D^BK^{(l)}_{BC}-\mathcal D _AK^{(l)}+\omega ^BK^{(l)}_{BC}-\omega _AK^{(l)}\bigg)\nn
-\beta\bigg(\mathcal D ^BK^{(n)}_{BC}-\mathcal D _AK^{(n)}+\omega ^BK^{(n)}_{BC}-\omega _AK^{(n)}\bigg)+g(n,C(X^T,\partial_A)l~),
\end{gather}
where the definition of Weyl tensor,
\begin{gather}
g\bigg(\partial_\mathfrak a,C(\partial_\mathfrak c,\partial_\mathfrak d)\partial_\mathfrak b\bigg)=g\bigg(\partial_\mathfrak a,R(\partial_\mathfrak c,\partial_\mathfrak d)\partial_\mathfrak b\bigg)-\bigg[g(\partial_\mathfrak a,\partial_{[\mathfrak c})R(\partial_{\mathfrak d]},\partial_\mathfrak b)-g(\partial_\mathfrak b,\partial_{[\mathfrak c})R(\partial_{\mathfrak d]},\partial_\mathfrak a)\bigg]\nn
+\frac{1}{3}Rg(\partial_\mathfrak a,\partial_{[\mathfrak c})g(\partial_{\mathfrak d]},\partial_\mathfrak b)\label{Weyl}
\end{gather}
has been used. A general expression for evolution along a vector of the form $X=\alpha l-\beta n$ is given in \cite{Booth:2006bn}. But here we have also considered the possiblity where the evolution vector might have a tangent component.
\section{Evolution of transverse extrinsic curvature}\label{EVTE}
Recall that $\lie_ng(\partial_A,\partial_B)=-2g\big(n,K(\partial_A,\partial_B)\big)$. Hence to calculate the metric to subleading order in $r$ one must know the transverse extrinsic curvature. This requires knowing the evolution equation for $g\big(n,K(\partial_A,\partial_B)\big)$ in the direction of $l$. Let us start by recalling the covariant derivative of the extrinsic curvature along an arbitrary normal direction $N$ such that $[N,\partial_A]=0$.
\begin{gather}
\nabla_NK(\partial_A,\partial_B)=\bigg(R(N^\perp,\partial_A)\partial_B\bigg)^\perp+\nabla^\perp_{\partial_A}\nabla^\perp_{\partial_B}N^\perp-\nabla^\perp_{\big(\nabla_{\partial_A}\partial_B\big)^T} N^\perp\nn
-K\big(\partial_A,W_{N^\perp}(\partial_B)\big)-g\bigg(K(\partial_A,\partial_B),\nabla^\perp_{\partial_C}N^\perp\bigg)\tilde q^{CD}\partial_D\nn
+\nabla_{N^T}K(\partial_A,\partial_B)-K([N^T,\partial_A],\partial_B)-K(\partial_A,[N^T,\partial_B])
\end{gather}
Now recall that,
\begin{gather}
g\big(n,\nabla^\perp_{\partial_B}~l\big)=\omega_B~l,~~g\big(n,\nabla^\perp_{\partial_A}\nabla^\perp_{\partial_B}~l\big)=-\partial_A\omega_B-\omega_A\omega_B,~~g\bigg(n,\nabla^\perp_{\big(\nabla_{\partial_A}\partial_B\big)^T}~l\bigg)=-\gamma^C_{AB}~\omega_C~l
\end{gather}
Using these equations it follows that,
\begin{gather}
\lie_{l}g\big(n,K(\partial_A,\partial_B)\big)
=-\kappa~g\big(n,K(\partial_A,\partial_B)\big)+g\big(R(l,\partial_A)\partial_B,n\big)-\mathcal D_A\omega_B-\omega_A\omega_B\nn
-g\big(n,K(\partial_A,\partial_C)\big)\tilde q^{CD}g\big(l,K(\partial_B,\partial_D)\big)\nn
\end{gather}
Interchanging $A$ and $B$ and adding, one has,
\begin{gather}
2\lie_{l}g\big(n,K(\partial_A,\partial_B)\big)=-2\kappa~g\big(n,K(\partial_A,\partial_B)\big)+g\big(R(l,\partial_A)\partial_B,n\big)+g\big(R(l,\partial_B)\partial_A,n\big)\nn
-2\mathcal D_{(A}\omega_{B)}-2\omega_{(A}\omega_{B)}-2g\big(n,K(\partial_C,\partial_{(A|})\big)\tilde q^{CD}g\big(l,K(\partial_{|B)},\partial_D)\big)\nn
\end{gather}
Now, using the Gauss' equation one has,
\begin{gather}
\lie_{l}g\big(n,K(\partial_A,\partial_B)\big)=-\kappa~g\big(n,K(\partial_A,\partial_B)\big)+\frac{1}{2}~^2\mathcal R(\partial_A,\partial_B)-\frac{1}{2}R(\partial_A,\partial_B)\nn
-\frac{1}{2}g(K(\partial_A,\partial_B),K)+\frac{1}{2}g(K(\partial_A,\partial_C),K(\partial_B,\partial_D))\tilde q^{CD}\nn
-\mathcal D_{(A}\omega_{B)}-\omega_{(A}\omega_{B)}-g\big(n,K(\partial_C,\partial_{(A|})\big)\tilde q^{CD}g\big(l,K(\partial_{|B)},\partial_D)\big),
\end{gather}
where $K:=\tilde q^{AB}K(\partial_A,\partial_B)$. In our case where the evolution is along a shear free and expansion free null geodesic, this equation simplies to,
\begin{gather}
\lie_{l}g\big(n,K(\partial_A,\partial_B)\big)=-\kappa~g\big(n,K(\partial_A,\partial_B)\big)+\frac{1}{2}~^2\mathcal R(\partial_A,\partial_B)-\frac{1}{2}R(\partial_A,\partial_B)-\mathcal D_{(A}\omega_{B)}-\omega_{(A}\omega_{B)}.\label{ETEC}
\end{gather}
\section{Calculation of $\nabla_ln$}
Let us assume that $\nabla_ln$ is of the form,
\begin{gather}
\nabla_ln=A~l+B~n+\chi^A\partial_A
\end{gather}
Contracting both sides by $n$ one has,
\begin{gather}
g(n,\nabla_ln)=-A=0,
\end{gather}
since $n$ is null throughout. Let us now contract both sides with $l$.
\begin{gather}
g(l,\nabla_ln)=-g(\nabla_l~l,n)=\kappa=-B
\end{gather}
Taking an inner product with $\partial_B$ one has,
\begin{gather}
g(\partial_B,\nabla_ln)=-g(\nabla_{\partial_B}l,n)=\omega_B=q_{AB}\chi^A
\end{gather}
Therefore we have
\begin{gather}
\nabla_ln=-\kappa n+\omega_Aq^{AB}\partial_B
\end{gather}
\bibliographystyle{ieeetr}
\bibliography{ref}

\end{document}